\documentclass[dvips]{article}
\usepackage{supertabular,lscape,epsfig}
\usepackage{amssymb}
\usepackage{amsmath}

\DeclareSymbolFont{ppa}{OT1}{ppl}{m}{it}
\DeclareMathSymbol{\vv}{\mathalpha}{ppa}{'166}

\begin{document}

\newcommand{\dd}{\,{\rm d}}
\newcommand{\ie}{{\it i.e.},\,}
\newcommand{\etal}{{\it et al.\ }}
\newcommand{\eg}{{\it e.g.},\,}
\newcommand{\cf}{{\it cf.\ }}
\newcommand{\vs}{{\it vs.\ }}
\newcommand{\zdot}{\makebox[0pt][l]{.}}
\newcommand{\up}[1]{\ifmmode^{\rm #1}\else$^{\rm #1}$\fi}
\newcommand{\dn}[1]{\ifmmode_{\rm #1}\else$_{\rm #1}$\fi}
\newcommand{\upd}{\up{d}}
\newcommand{\uph}{\up{h}}
\newcommand{\upm}{\up{m}}
\newcommand{\ups}{\up{s}}
\newcommand{\arcd}{\ifmmode^{\circ}\else$^{\circ}$\fi}
\newcommand{\arcm}{\ifmmode{'}\else$'$\fi}
\newcommand{\arcs}{\ifmmode{''}\else$''$\fi}
\newcommand{\MS}{{\rm M}\ifmmode_{\odot}\else$_{\odot}$\fi}
\newcommand{\RS}{{\rm R}\ifmmode_{\odot}\else$_{\odot}$\fi}
\newcommand{\LS}{{\rm L}\ifmmode_{\odot}\else$_{\odot}$\fi}

\newcommand{\Abstract}[2]{{\footnotesize\begin{center}ABSTRACT\end{center}
\vspace{1mm}\par#1\par
\noindent
{~}{\it #2}}}

\newcommand{\TabCap}[2]{\begin{center}\parbox[t]{#1}{\begin{center}
  \small {\spaceskip 2pt plus 1pt minus 1pt T a b l e}
  \refstepcounter{table}\thetable \\[2mm]
  \footnotesize #2 \end{center}}\end{center}}

\newcommand{\TableSep}[2]{\begin{table}[p]\vspace{#1}
\TabCap{#2}\end{table}}

\newcommand{\FigCap}[1]{\footnotesize\par\noindent Fig.\  %
  \refstepcounter{figure}\thefigure. #1\par}

\newcommand{\TableFont}{\footnotesize}
\newcommand{\TableFontIt}{\ttit}
\newcommand{\SetTableFont}[1]{\renewcommand{\TableFont}{#1}}

\newcommand{\MakeTable}[4]{\begin{table}[htb]\TabCap{#2}{#3}
  \begin{center} \TableFont \begin{tabular}{#1} #4 
  \end{tabular}\end{center}\end{table}}

\newcommand{\MakeTableSep}[4]{\begin{table}[p]\TabCap{#2}{#3}
  \begin{center} \TableFont \begin{tabular}{#1} #4 
  \end{tabular}\end{center}\end{table}}

\newenvironment{references}%
{
\footnotesize \frenchspacing
\renewcommand{\thesection}{}
\renewcommand{\in}{{\rm in }}
\renewcommand{\AA}{Astron.\ Astrophys.}
\newcommand{\AAS}{Astron.~Astrophys.~Suppl.~Ser.}
\newcommand{\ApJ}{Astrophys.\ J.}
\newcommand{\ApJS}{Astrophys.\ J.~Suppl.~Ser.}
\newcommand{\ApJL}{Astrophys.\ J.~Letters}
\newcommand{\AJ}{Astron.\ J.}
\newcommand{\IBVS}{IBVS}
\newcommand{\PASP}{P.A.S.P.}
\newcommand{\Acta}{Acta Astron.}
\newcommand{\MNRAS}{MNRAS}
\renewcommand{\and}{{\rm and }}
\section{{\rm REFERENCES}}
\sloppy \hyphenpenalty10000
\begin{list}{}{\leftmargin1cm\listparindent-1cm
\itemindent\listparindent\parsep0pt\itemsep0pt}}%
{\end{list}\vspace{2mm}}

\def\TYLDA{~}
\newlength{\DW}
\settowidth{\DW}{0}
\newcommand{\dw}{\hspace{\DW}}

\newcommand{\refitem}[5]{\item[]{#1} #2%
\def\REFARG{#3}\ifx\REFARG\TYLDA\else, {\it#3}\fi
\def\REFARG{#4}\ifx\REFARG\TYLDA\else, {\bf#4}\fi
\def\REFARG{#5}\ifx\REFARG\TYLDA\else, {#5}\fi.}

\newcommand{\Section}[1]{\section{#1}}
\newcommand{\Subsection}[1]{\subsection{#1}}
\newcommand{\Acknow}[1]{\par\vspace{5mm}{\bf Acknowledgements.} #1}
\pagestyle{myheadings}

\newfont{\bb}{ptmbi8t at 12pt}
\newcommand{\xrule}{\rule{0pt}{2.5ex}}
\newcommand{\xxrule}{\rule[-1.8ex]{0pt}{4.5ex}}
\def\thefootnote{\fnsymbol{footnote}}
\begin{center}
{\Large\bf Detached Eclipsing Binaries for Parallax Measurements}
\vskip1cm
Irena~~~~S~e~m~e~n~i~u~k
\vskip3mm
Warsaw University Observatory, Al.~Ujazdowskie~4, 00-478~Warszawa, Poland\\
e-mail: is@astrouw.edu.pl
\end{center}

\Abstract{A list of 50 bright Detached Eclipsing Binaries proposed 
for precise parallax measurements for the FAME is presented. The
eclipsing binaries, with precise distances and with photometric
and double-line spectroscopic orbits determined, are essential for
a full calibration of relations that could be used for very precise
distance determination of globular clusters, LMC, SMC, and perhaps
also M31 and M33. 
}

It is expected that since 2004 the Full-sky Astrometric Mapping Explorer
(FAME) will be operating from space (Horner et al. 1999). The spacecraft
will determine, in particular, parallaxes with accuracy to 50
microarcseconds for stars between 5th and 9th visual magnitudes.
In this connection the author was asked by Prof. Bohdan Paczy{\'n}ski 
to prepare a list of 50 bright detached eclipsing binaries,
that would be - after having precisely measured distances -
particularly good candidates for empirical calibration of the distance
determination based on detached eclipsing binaries. This is a very old,
and potentially excellent method; a large number of references, as well
as the description of the method is provided by Kruszewski and Semeniuk
(1999). A comparison between the distances based on eclipsing binaries
(Lacy 1979) and Hipparcos parallaxes demonstrated that the method is
indeed promissing (Semeniuk 2000), but there is plenty of room for an
improvement. 

The binary stars proposed for FAME program are expected to be subjects
of a special treatement during the parallax measurements.
 
The proposed stars should fulfill the following imposed criteria:

1. They should be fainter than 5 and brighter than 9 visual magnitude.

2. Their distances should be smaller than 1 kps.

3. The primary and secondary minima of a star should have comparable depths.

4. Their spectral types should be from the range B0 - K9.

5. They should have no characteristics of chemical peculiarity in 
   their spectra, \ie Am or Fm stars should be excluded.

6. Their components should have no intrinsic variability. 

It would also be desirable from the point of view of a surface
brightness-color relation that the spectral types be approximately 
uniformly cast by stars, with each spectral type represented by 
about 10 stars. 

Table 1 contains proposed detached eclipsing binaries.
They were selected generally from the {\it Hipparcos Variability Annex} 
\ie the Vol. 11 of the {\it Hipparcos Catalogue} (ESA 1997). This volume,
in its Section "Periodic Variables", contains 410 stars classified as
eclipsing binaries of the variability type EA.

The stars in Table 1 are arranged according to the spectral types.
The columns of the table are generally self-explanatory. Comments must
only be given to some of them. The depth of minima in Columns 4 and 5
are generally in $V$ magnitude. The letters y or b, that follow some
of the depth values, denote, that they are in $y$ or $b$ magnitudes of 
the Str\"{o}mgren four--color ${\it ubvy}$ system. The parallaxes in
Column 8 were taken from the {\it Hipparcos Catalogue}, when their
standard errors were less than 25\% and we had no distance
determination from the analysis of photometric and spectroscopic 
orbits. For stars with such distances, and with the
Hipparcos parallax error greater than 20\%, we preferred to place
them in Column 8 instead of Hipparcos parallaxes. The parallax values
and their errors, in such cases, were taken from available literature. 
In the case when no error was given in the literature we estimated it assuming
the value for the relative error of parallax equal to 0.07. This value
was obtained as the mean from the relative errors published in the literature.
The parallaxes obtained from the photometric and spectroscopic elements are 
followed by asterisk. For the star V335 Ser with no parallax value in 
the literature we have estimated it using the spectral type - luminosity
calibration (Lang 1992). It is denoted by a cross following the parallax 
value. Columns 11 to 14 (on the second page of the table) contain relative
radii ${a_1}$ and ${a_2}$ of the components of the system and their
errors ${\sigma_{a1}}$, ${\sigma_{a2}}$ and Columns 15 to 18 their radial 
velocities ${K_1}$ and ${K_2}$ and their errors ${\sigma_{K1}}$ and 
${\sigma_{K2}}$, respectively. We can see that the accuracy both of the 
relative radii and the radial velocity amplitudes is generally on the 
level of the last figure of their values. Table 1 contains 35 systems with
determined elements
both of photometric and spectroscopic orbits. For the remaining 15
systems the author could not find in available literature any information
concerning such orbits except for HP~Dra and V2080~Cyg whose elements of
spectroscopic orbits have recently been determined.   
   
As we can see Table 1 is dominated by stars with the spectral types 
B, A and F. There are only a few stars with G spectral type and there 
are no systems with K type components except for one component of AI Phe.
The reason for that is that we could hardly ever find, among the G and K 
spectral type binaries, stars with no signs of chromospheric activity.       
For the systems with G type components, that are placed in Table 1,    
we have no indications, at the moment, that they are chromospherically 
active. However, it is not excluded that future thorough photometric 
observations may indicate an RS~CVn~type variability in their light curves.
This concerns particularly the two stars with short orbital periods \ie 
KR~Aqr and UW~LMi, as such variability in this spectral type depends 
strongly on the orbital period and is much more likely for systems
with short orbital periods than for the stars with longer periods.  

\Acknow{This paper was undertaken on suggestion of Professor Bohdan 
Paczy{\'n}ski. The author is greatly indebted to him for managing
discussions and for reading and commenting on the manuscript. Special
thanks are due to Prof.~Andrzej Kru\-szew\-ski for continuous interest 
and many stimulating discussions. Professors Kazimierz St\c epie{\'n} and
Wojciech Dziembowski are acknowledged for helpful discussions on
the chromospheric activity of the late spectral type stars. This work
would not be possible without use of the {\it Hipparcos Catalogue}.}

\end{document}